\documentclass[aps,pra,reprint,superscriptaddress,nofootinbib,longbibliography,floatfix]{revtex4-2} 

\usepackage[T1]{fontenc}
\usepackage[utf8]{inputenc}
\usepackage{amsmath,amssymb,amsfonts,mathtools,bm,braket,physics}
\usepackage{graphicx}
\usepackage{xcolor}
\usepackage{silence}
\usepackage{hyperref}
\usepackage{booktabs}
\usepackage{multirow}
\usepackage{enumitem}
\usepackage{microtype}
\usepackage{ragged2e}
\usepackage[nameinlink,noabbrev]{cleveref}
\hypersetup{%
    colorlinks=true,
    linkcolor=blue,
    citecolor=blue,
    urlcolor=blue
}

\makeatletter
\def\@bibdataout@aps{%
  \immediate\write\@bibdataout{%
    @CONTROL{apsrev42Control,author="48",editor="1",pages="0",title="0",year="1"}%
  }%
  \if@filesw%
    \immediate\write\@auxout{\string\citation{apsrev42Control}}%
  \fi
}
\makeatother

\begin{document}

\title{Practical Fidelity Estimation for Noisy Quantum Devices Using Direct and Survival-Probability-Based Protocols}

\author{Ahmed Aman}
\affiliation{Indian Institute of Science Education and Research Mohali, Punjab 140306, India}
\email{ahm3daman@gmail.com}

\author{Kaustav Chatterjee}
\affiliation{Department of Physics, Quantum Physics and Information Technology, Technical University of Denmark, 2800 Kgs. Lyngby, Denmark}

\author{Younes Javanmard}
\affiliation{Institut f\"ur Theoretische Physik, Leibniz Universit\"at Hannover, Appelstra{\ss}e 2, 30167 Hannover, Germany}

\author{Tanmoy Pandit}
\affiliation{VTT Technical Research Centre of Finland, Tietotie 3, 02150 Espoo, Finland}
\email{tanmoy.pandit@vtt.fi}

\begin{abstract}
Reliable characterization of quantum states and quantum processes is central to the development of scalable quantum computing. Standard characterization techniques such as full quantum state tomography become rapidly impractical as system size increases. In this work we present two software-oriented fidelity estimation protocols that avoid full tomographic reconstruction while remaining experimentally relevant for noisy intermediate-scale quantum devices. The first protocol is direct fidelity estimation (DFE), where the fidelity between a target state and an experimentally prepared state is estimated from importance-sampled Pauli measurements. The second protocol is a survival-probability-based method that reconstructs fidelity from experimentally accessible moments of a noisy evolution and its approximate inverse, and is intrinsically robust against state-preparation-and-measurement (SPAM) errors. We discuss the theoretical basis of both approaches, analyze their noise sensitivity, and present numerical evidence on representative multiqubit states under depolarizing, dephasing, and amplitude-damping channels. For DFE we show that attenuation-aware correction improves robustness against measurement noise. For the survival-probability approach we show that an $L_2$-optimized expansion substantially outperforms naive Taylor truncation in high-noise regimes. 
\end{abstract}

\maketitle

\section{Introduction}
Benchmarking quantum devices requires estimators that are both experimentally feasible and computationally scalable. Although complete tomography provides a conceptually direct route to state or process reconstruction, its measurement and post-processing costs scale unfavorably with system size. This motivates the development of task-specific estimators that target operational figures of merit such as fidelity without reconstructing the full density matrix ~\cite{huang2020predicting,elben2020mixedstate,elben2023toolbox,gross2010compressed} .

Among the most influential examples is direct fidelity estimation (DFE), introduced by Flammia and Liu, which estimates the overlap between an unknown state and a known target state using importance sampling over Pauli observables~\cite{flammia2011direct}. Practical fidelity estimation without full tomography has also been studied from a broader device-characterization perspective~\cite{dasilva2011practical}. More recently, improved sampling strategies based on grouping Pauli operators have been shown to reduce copy complexity in favorable settings~\cite{barbera2025sampling}, while generic-state extensions broaden the applicability of DFE beyond special target families~\cite{vairogs2024generic}. Alternative approaches include fidelity estimators with nearly minimax confidence intervals and data-driven experimental fidelity learning~\cite{seshadri2024versatile,qin2024experimental}.

Related low-overhead diagnostics exploit periodic driving to identify departures from unitary evolution using repeated measurements of a single observable. These methods can detect incoherent errors without full tomographic reconstruction and provide a complementary dynamics-based perspective on scalable device characterization~\cite{green2022probing}. Repeated sequential measurements have also been used to characterize incoherent noise while reducing contamination from coherent control errors, including robust estimation of the dephasing time in the presence of Markovian and non-Markovian noise~\cite{penshin2024robust}.

In this paper we combine two complementary fidelity-estimation routes into a unified benchmarking framework and compare their advantages and limitations across different benchmarking scenarios. The first is DFE, which is efficient when the target state's Pauli structure is known and experimentally measurable. The second is a survival-probability-based framework that characterizes the cumulative effect of a noisy quantum evolution represented by a superoperator $\mathcal K$. Rather than estimating fidelity directly from expectation values of many observables, this approach exploits return probabilities associated with a forward--inverse evolution cycle. The central object is the positive-semidefinite operator $ \mathcal Q = \mathcal K^\dagger \mathcal K$ which quantifies deviations from ideal unitary evolution. Moments of $\mathcal Q$ can be accessed experimentally through repeated survival-probability measurements and subsequently combined to reconstruct an adjoint-cycle fidelity proxy. When the implemented inverse operation realizes the adjoint evolution $\mathcal K^\dagger$, this proxy approaches the physical fidelity associated with the noisy process. More generally, however, it should be viewed as a diagnostic quantity that captures information about the accumulated noise rather than the exact state fidelity. Return-probability measurements can offer practical advantages under state-preparation-and-measurement (SPAM) errors, although their robustness ultimately depends on the quality of the experimental inverse implementation.

The main contributions of this paper are as follows. First, we present a clear software-level formulation of DFE suitable for numerical and experimental implementation. Second, we analyze the effect of measurement noise and explain how attenuation-factor corrections improve the fidelity estimates. Third, we formulate a survival-probability reconstruction of the adjoint-cycle fidelity proxy, make the role of its expansion coefficients transparent, and show why an $L_2$-norm optimization is preferable to a naive Taylor expansion in strong-noise regimes.

\begin{figure*}[t]
    \centering
    \includegraphics[width=0.9\textwidth]{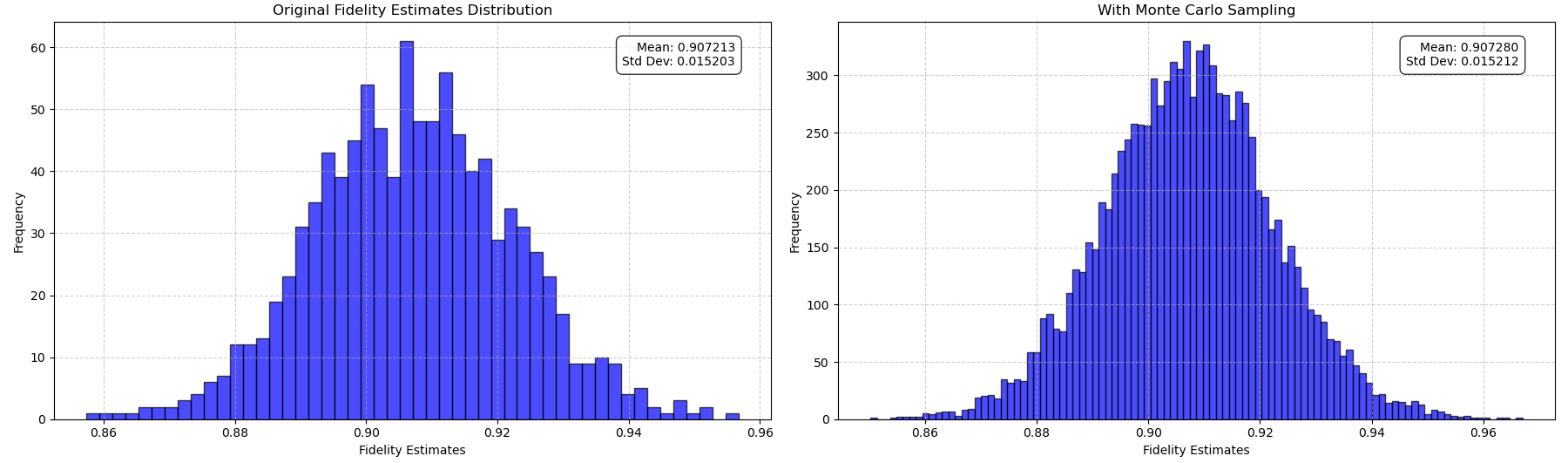}
    \caption{Representative DFE results: Distribution of the fidelity estimates obtained from direct simulations and from the Monte Carlo extension. This distribution is based on a four-qubit Haar-random state that underwent a $10\%$ depolarizing noise. The true calculated fidelity came out to be 0.90626, and the average from the plot was 0.907213. The plot on the left is a simulation of 1000 raw samples, whereas the plot on the right is with the Monte Carlo sampling from 1000 raw initial samples that was extended to 9000 samples. The absolute mean difference between the original estimation distribution and the distribution with the Monte Carlo sampling is 0.00102 for this particular case.}\label{fig:dfe_results}
\end{figure*}

\section{Theoretical Framework}
\subsection{Direct fidelity estimation}
\subsubsection{Fidelity under depolarizing noise}
\begin{figure}[t!]
    \centering
    \includegraphics[width=0.88\columnwidth]{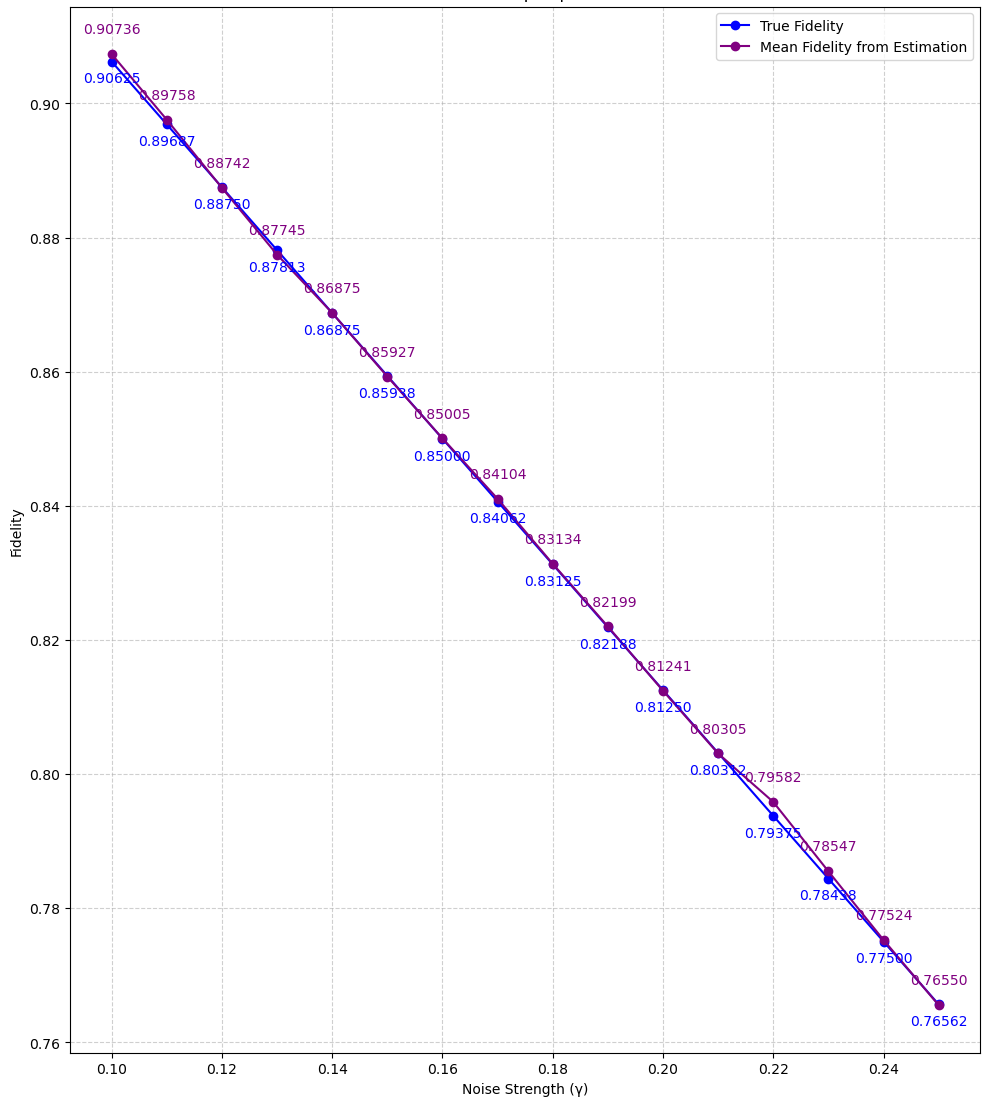}
    \caption{True fidelity and mean estimated fidelity as functions of noise strength for a four-qubit Haar-random state under global depolarizing noise. The true-fidelity reference follows \cref{eq:depolarizing-fidelity} with $d=16$, and the simulated DFE means provide a numerical check of this result.}
    \label{fig:dfe_noise_sweep}
\end{figure}

For a pure target state $\rho=\ketbra{\psi}{\psi}$ in dimension $d$, global depolarizing noise of strength $p$ prepares
\begin{equation}
\sigma=(1-p)\rho+p\frac{\mathbb{I}}{d}.
\end{equation}
Since $\Tr(\rho^2)=\Tr(\rho)=1$, the fidelity used by DFE is
\begin{align}
F(\rho,\sigma)
&=\Tr\!\left[\rho\left((1-p)\rho+p\frac{\mathbb{I}}{d}\right)\right]\\
&=1-p\left(1-\frac{1}{d}\right).
\label{eq:depolarizing-fidelity}
\end{align}
Thus, for $n$ qubits, $F=1-p(1-2^{-n})$, independently of the particular pure target state or preceding ideal unitary. The channel properties underlying this result are summarized in Appendix~\ref{app:depolarizing-fidelity}. In the four-qubit noise-sweep simulation in \cref{fig:dfe_noise_sweep}, the blue curve is calculated using Eq.~\eqref{eq:depolarizing-fidelity}.

\subsubsection{Protocol}
Let $\rho$ be a pure target state in a Hilbert space $\mathcal{H}$ of dimension $d=2^n$, and let $\sigma$ denote the experimentally prepared state. We assume that one can measure $n$-qubit Pauli observables, defined as tensor products of single-qubit Pauli operators, without imposing additional structure or symmetry on $\rho$. The fidelity of interest is
\begin{equation}
F(\rho,\sigma)=\Tr(\rho\sigma).
\end{equation}
Expanding both states in the normalized Pauli basis $\{W_k\}_{k=1}^{4^n}$, we define the characteristic functions
\begin{equation}
\mathcal{X}_{\rho}(k)=\frac{\Tr(\rho W_k)}{\sqrt{d}},
\qquad
\mathcal{X}_{\sigma}(k)=\frac{\Tr(\sigma W_k)}{\sqrt{d}},
\end{equation}
so that
\begin{equation}
F(\rho,\sigma)=\sum_k \mathcal{X}_{\rho}(k)\mathcal{X}_{\sigma}(k).
\end{equation}
The estimation uses Pauli operators selected according to the importance-weighted distribution
\begin{equation}
p(k)=\mathcal{X}_{\rho}(k)^2,
\end{equation}
which is valid for pure $\rho$ because $\sum_k \mathcal{X}_{\rho}(k)^2=1$.

For $k$ drawn from $p(k)$, define the random variable
\begin{equation}
X=\frac{\mathcal{X}_{\sigma}(k)}{\mathcal{X}_{\rho}(k)}.
\end{equation}
Its expectation value is $\mathbb{E}[X]=\sum_k p(k)X=\Tr(\rho\sigma)$.

The DFE protocol is as follows~\cite{flammia2011direct}:

\begin{enumerate}
        \item Generate $\ell$ independent and identically distributed samples $k_1,\ldots,k_\ell$ from $p(k)$, where
        \begin{equation}
            \ell := \lceil (\epsilon^2 \delta)^{-1}\rceil
        \end{equation}
        \item Measure each observable $W_{k_i}$ a number of $m_{i}$ times, where $m_{i}$ is defined as
        \begin{equation}\label{eq:copies-original}
            m_i = \left\lceil \frac{2}{[\mathcal{X}_\rho(k_i)]^2 d\ell\epsilon^2}\ln\!\left(\frac{2}{\delta}\right) \right\rceil.
        \end{equation}
        Thus, $m_i$ depends on the sampled index $k_i$.
        \item Measure $W_{k_i}$ on $m_i$ copies of $\sigma$, obtaining outcomes $A_{ij}\in\{-1,1\}$. These outcomes provide an empirical estimate of $\Tr(\sigma W_{k_i})$ and define
        \begin{equation}
            \tilde{X}_i = \frac{1}{m_i\sqrt{d}\,\mathcal{X}_\rho(k_i)} \sum_{j=1}^{m_i} A_{ij}.
        \end{equation}
        \item Calculate
        \begin{equation}
            \tilde{Y} = \frac{1}{\ell} \sum^\ell_{i=1} \tilde{X}_i
        \end{equation}
        \item Return $\tilde{Y}$ as the final fidelity estimator.
    \end{enumerate}

The fidelity estimate $\tilde{Y}$ is an unbiased estimator of $F(\rho,\sigma)$ with accuracy $2\epsilon$ and confidence level $1-2\delta$. The protocol guarantees that the expected number of state preparations is bounded and scales at most linearly with the Hilbert-space dimension.

\subsubsection{Readout-noise model}
DFE is sensitive to measurement noise because it relies directly on estimated Pauli expectation values. If each measured qubit is independently flipped with probability $\alpha$, then an ideal expectation value of a Pauli string $W_k$ is attenuated according to
\begin{equation}
        \expval{W_k}_{\mathrm{meas}}
=\beta_k\expval{W_k}_{\mathrm{ideal}},
\qquad \beta_k=(1-2\alpha)^{w(k)},
\end{equation}
Here $w(k)$ is the number of nonidentity single-qubit factors in the Pauli string $W_k$. Independent bit flips on each qubit therefore attenuate each nonidentity factor, and the attenuation multiplies across the string. For a weight-one string, or for a single effective binary outcome channel, this reduces to $\beta=1-2\alpha$. Consequently, the attenuation induces an observable-dependent bias in the raw DFE estimate by suppressing nonidentity contributions. A simple correction is to divide the measured expectation value by $\beta_k$, or more generally by a calibrated observable-dependent attenuation coefficient if the readout model is nonuniform across qubits~\cite{bravyi2021mitigating}. This correction requires nonzero attenuation and can amplify statistical noise when $|\beta_k|$ is small. In the numerical regime considered here, the effective binary-outcome correction substantially improves the agreement between the mean estimated fidelity and the true fidelity, even for measurement error rates up to roughly $5\%$.

\subsection{Survival-probability-based fidelity estimation}
\subsubsection{Operational protocol}
We now consider a different strategy that avoids explicit Pauli reconstruction. Let $\rho_0=\ketbra{\psi}{\psi}$ be a known initial state. We work in Liouville space, where an operator $A$ is represented by the vector $\lvert A\rangle\rangle$ and the inner product is the Hilbert--Schmidt inner product
\begin{equation}
\langle\langle A\vert B\rangle\rangle
=\Tr\!\left(A^{\dagger}B\right).
\label{eq:liouville-inner-product}
\end{equation}
Let $\mathcal{U}$ denote the Liouville-space superoperator associated with the intended unitary evolution, $\mathcal{U}(\rho)=U\rho U^{\dagger}$. The implemented forward evolution is modeled by the superoperator
\begin{equation}
\mathcal{K}=\mathcal{U}\exp\!\left(\sum_i\Omega_i\right),
\end{equation}
where the $\Omega_i$ are Liouville-space generators describing deviations from the ideal evolution. Define the interaction-picture error map $\mathcal{E}=\exp(\sum_i\Omega_i)$, so $\mathcal{K}=\mathcal{U}\mathcal{E}$. For the ideal output $\lvert\rho_f^{\mathrm{ID}}\rangle\rangle=\mathcal{U}\lvert\rho_0\rangle\rangle$ and noisy output $\lvert\rho_f\rangle\rangle=\mathcal{K}\lvert\rho_0\rangle\rangle$, the physical state fidelity is
\begin{equation}
F_{\mathrm{phys}}=\langle\langle\rho_f^{\mathrm{ID}}\vert\rho_f\rangle\rangle
=\langle\langle\rho_0\vert\mathcal{E}\vert\rho_0\rangle\rangle.
\end{equation}
The polynomial reconstruction below instead targets the adjoint-cycle fidelity proxy
\begin{equation}
\mathcal{Q}=\mathcal{K}^{\dagger}\mathcal{K}=\mathcal{E}^{\dagger}\mathcal{E},
\qquad
F_{\mathcal{Q}}=\langle\langle\rho_0\vert\sqrt{\mathcal{Q}}\vert\rho_0\rangle\rangle.
\label{eq:adjoint-cycle-proxy}
\end{equation}
In general, $F_{\mathcal{Q}}\neq F_{\mathrm{phys}}$. They coincide if $\mathcal{E}$ is positive semidefinite, so that $\sqrt{\mathcal{E}^{\dagger}\mathcal{E}}=\mathcal{E}$.

Let $\mathcal{K}_I$ denote the experimentally implemented approximate inverse, which ideally reproduces $\mathcal{U}^{\dagger}$. A single forward--inverse cycle is therefore described by
\begin{equation}
\mathcal{C}=\mathcal{K}_I\mathcal{K}.
\end{equation}
The experimentally accessible survival probabilities are
\begin{equation}
R_m=\langle\langle\rho_0\vert\mathcal{C}^{m}
\vert\rho_0\rangle\rangle
=\langle\langle\rho_0\vert
(\mathcal{K}_I\mathcal{K})^{m}
\vert\rho_0\rangle\rangle.
\label{eq:survival-probability}
\end{equation}
Thus, $R_m$ is obtained by applying $m$ forward--inverse cycles and measuring the return probability to the initial state. These are the moments of $\mathcal{Q}$ only if the implemented inverse satisfies $\mathcal{K}_I=\mathcal{K}^{\dagger}$; because $\mathcal{K}^{\dagger}$ need not be trace-preserving, this equality is a condition on the implemented map, not a generic physical-inversion guarantee. A generic physical approximate inverse instead yields moments of $\mathcal{C}$ and does not, by itself, reconstruct $F_{\mathcal{Q}}$. The physical pulse-inverse construction and spectral validity conditions are derived in \cref{app:survival_derivations}. Subject to the spectrum of $\mathcal{Q}$ lying in the approximation interval, the proxy in \cref{eq:adjoint-cycle-proxy} is reconstructed through an expansion of the form
\begin{equation}
F_{\mathcal{Q}} \approx c_0R_0+c_1R_1+\cdots+c_MR_M.
\label{eq:survival_expansion}
\end{equation}
Return-probability measurements may reduce some SPAM sensitivities relative to DFE, but this advantage depends on the implementation and is not quantified by the exact-moment simulations reported here.

\begin{figure}[t!]
    \centering
    \includegraphics[width=\columnwidth]{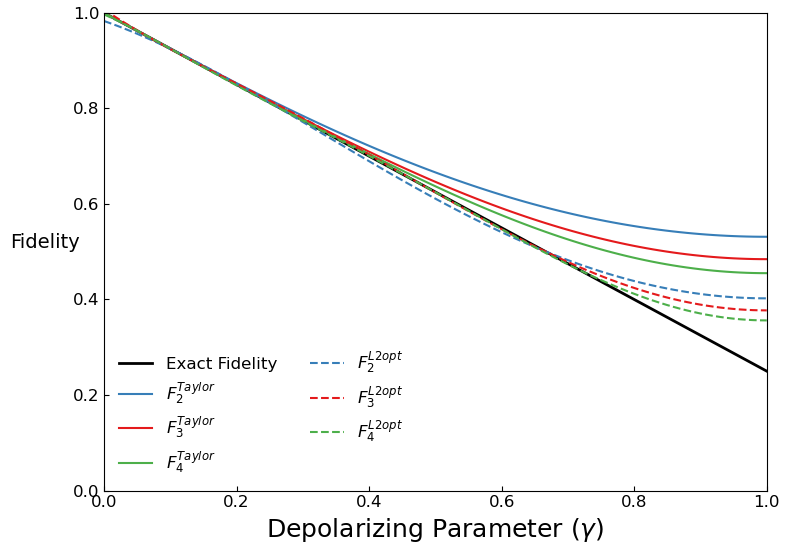}
\caption{Survival-probability reconstruction of the adjoint-cycle fidelity proxy. The dotted curve corresponds to the reconstruction with $L_2$-optimized coefficients, which improves the high-noise behaviour relative to a standard Taylor-based approximation. This particular plot is based on a simulation using a 2-qubit Haar random state undergoing depolarizing channel.}
    \label{fig:survival}
\end{figure}

\subsubsection{Optimized polynomial reconstruction}
The accuracy of \cref{eq:survival_expansion} depends critically on the choice of the coefficients $c_m$. A straightforward approach is to derive them from a low-order Taylor expansion around the weak-noise limit. However, such coefficients are only locally optimal and can fail badly when the effective action of the noise is not perturbatively small. This is precisely the regime where a more globally optimized approximation becomes useful.

We therefore choose the coefficients by minimizing an $L_2$ approximation error over a physically relevant interval $\lambda\in[g,1]$:
\begin{equation}
\epsilon_{L_2}^{M} = \int_{g}^{1}
\left(\sum_{m=0}^{M}a_m^{M}\lambda^m - \lambda^{1/2}\right)^2 d\lambda.
\label{eq:l2_error}
\end{equation}
The interval formulation assumes that the relevant cycle spectrum is real and contained in $[g,1]$. This condition holds directly for the positive adjoint cycle $\mathcal{Q}=\mathcal{K}^{\dagger}\mathcal{K}$ when its largest eigenvalue is at most unity, or after an appropriate spectral rescaling. A physical pulse-inverse cycle $\mathcal{C}=\mathcal{K}_I\mathcal{K}$ equals $\mathcal{Q}$ only if $\mathcal{K}_I=\mathcal{K}^{\dagger}$; otherwise positivity and a real spectrum must be verified separately, and the reconstructed quantity is not automatically $F_{\mathcal{Q}}$. A lower spectral bound and the associated validity conditions are given in \cref{app:spectral_bounds}.
The stationarity conditions
\begin{equation}
\frac{\delta \epsilon_{L_2}^{M}}{\delta a_m^{M}}=0
\end{equation}
lead to the linear system
\begin{equation}
\sum_{n=0}^{M}G_{mn}a_n^M=b_m,
\qquad m=0,\ldots,M,
\end{equation}
where
\begin{equation}
G_{mn}=\frac{1-g^{m+n+1}}{m+n+1},
\qquad
b_m=\frac{1-g^{m+3/2}}{m+3/2}.
\end{equation}
For the $L_2$-optimized estimator, the coefficients in \cref{eq:survival_expansion} are identified as
\begin{equation}
c_m=a_m^M,\qquad m=0,\ldots,M.
\end{equation}
These coefficients provide a polynomial approximation to $\lambda^{1/2}$ that is optimized over the interval of interest rather than around a single expansion point. In practice, this produces a markedly improved fidelity reconstruction in the strong-noise regime.

\subsubsection{Interpretation and SPAM robustness}
The operational advantage of the survival-probability method is that it uses repeated return-probability measurements rather than many distinct Pauli expectation values. Such measurements more stable under readout imperfections in some implementations and for details one can check the Appendix C of the \cite{green2022probing}. The reconstructed quantity is $F_{\mathcal{Q}}$ when the measured cycle equals $\mathcal{Q}$; for a general physical approximate inverse, the data instead describe the cycle $\mathcal{C}$ and require a separately justified reconstruction. Conceptually, DFE and survival-probability estimation are complementary: DFE is a state-structured protocol that leverages the Pauli representation of a known target, whereas the survival-probability method is dynamics-structured and leverages moments of an evolution cycle.

\section{Numerical Methods}
We considered representative families of multiqubit target states, including Haar-random states, GHZ states, and $W$ states. Noise was introduced through standard single- and multiqubit channels, including depolarizing, dephasing, and amplitude-damping processes. The simulations were implemented in Python using QuTiP-based state preparation and noisy evolution. Check the link given under Data Availability for the other plots.

\subsection{Direct fidelity estimation simulations}
The DFE calculations used four-qubit target states with $\epsilon=\delta=0.05$, corresponding to $\ell=8000$ importance-sampled Pauli indices per trial. Haar-random, GHZ, and $W$ targets were tested under global depolarizing noise and local phase-damping, amplitude-damping, and generalized amplitude-damping channels. Pauli outcomes were generated by binomial sampling, with the number of shots $m_i$ assigned according to \cref{eq:copies-original}. An effective binary readout-flip probability of $p_{\mathrm{flip}}=0.05$ was applied to each sampled Pauli outcome in the noise-strength sweep; this models a flip of the effective binary result, not independent bit flips on each qubit. The measured expectation values were divided by the calibrated attenuation factor $1-2p_{\mathrm{flip}}$. The noise strength was varied from $0.10$ to $0.25$ in increments of $0.01$, with 1000 independent DFE trials at each value, and the estimates were compared with the directly calculated overlap $\Tr(\rho\sigma)$.

The shot simulation was vectorized to reduce runtime. A second practical optimization concerned repeated Monte Carlo trials. Rather than simulating every shot configuration independently for the full ensemble size, we used a two-stage procedure: a smaller set of full simulations was first generated to identify an empirical distribution for the estimator, and then additional samples were drawn from the fitted distribution to extend the statistics. This gives a substantial speedup while preserving the overall structure of the estimator histogram, provided that the fitted distribution is validated against the directly simulated subset. The directly simulated and extended ensembles are shown separately in \cref{fig:dfe_results} so that the effect of this approximation remains visible. For the distribution comparison in \cref{fig:dfe_results}, 1000 directly simulated estimates were fitted to a candidate parametric distribution and supplemented with 9000 synthetic draws. The direct and extended ensembles are displayed separately because the latter increase the apparent sample size without adding independent quantum-measurement simulations. 

\subsection{Survival-probability simulations}
The survival-probability reconstruction was evaluated for a two-qubit Haar-random pure target state under global depolarizing noise, independent local amplitude damping, and collective amplitude damping. For each channel, the noise strength was sampled at 100 uniformly spaced points in $[0,1]$. The numerical implementation used the positive adjoint cycle $\mathcal{Q}=\mathcal{K}^{\dagger}\mathcal{K}$ and evaluated the exact moments
\begin{equation}
R_m=\langle\langle\rho_0\vert\mathcal{Q}^{m}\vert\rho_0\rangle\rangle,
\qquad m=0,\ldots,M.
\end{equation}
Reconstructions of orders $M=2,3,4$ were compared using the Taylor coefficients and the $L_2$-optimized coefficients obtained from \cref{eq:l2_error} with $g=0.05$. The reference quantity was the proxy $F_{\mathcal{Q}}$ defined in \cref{eq:adjoint-cycle-proxy}, evaluated by a direct matrix square root. Thus, these simulations benchmark the adjoint-cycle reconstruction; they represent moments of an experimental forward--inverse protocol only when $\mathcal{K}_I=\mathcal{K}^{\dagger}$. They equal physical state fidelity only when the interaction-picture error map $\mathcal{E}$ is positive semidefinite. The baseline curves use exact moments and therefore do not include finite-shot or explicit SPAM fluctuations. The depolarizing-noise example is shown in \cref{fig:survival}.

\section{Results and Discussion}
\subsection{Comparison of the two protocols}
The two protocols discussed here address different experimental constraints and should be viewed as complementary tools rather than competing replacements.

First, DFE is particularly natural when the target state is known classically and its relevant Pauli coefficients can be computed efficiently. In that case, the protocol offers a direct and statistically interpretable path to fidelity estimation with far fewer settings than full tomography. Its main drawback is sensitivity to measurement noise and, more broadly, the need to perform many distinct measurement bases when the target does not have a sparse or structured Pauli representation.

Second, the survival-probability method is attractive when one can access repeated return probabilities and wants a dynamics-based proxy without Pauli-level interpretability. Its polynomial reconstruction estimates $F_{\mathcal{Q}}$ only when the measured cycle realizes $\mathcal{Q}=\mathcal{K}^{\dagger}\mathcal{K}$ and the stated spectral assumptions hold; otherwise a different cycle-specific target must be justified. The expansion-coefficient viewpoint opens a path to optimization: rather than fixing coefficients from perturbation theory, one can tailor them to the expected spectral interval by solving a controlled approximation problem such as \cref{eq:l2_error}.

The survival-probability method is conceptually related to other repeated-measurement diagnostics~\cite{pandit2022bounds,green2022probing,penshin2024robust}, since these approaches extract noise information from sequential dynamical cycles. Their objectives are nevertheless different. Periodic-driving signatures diagnose violations of unitarity or periodicity~\cite{pandit2022bounds,green2022probing}, while robust sequential-measurement protocols isolate incoherent noise parameters from coherent control errors~\cite{penshin2024robust}. The present method instead reconstructs the adjoint-cycle fidelity proxy from a polynomial combination of return probabilities, subject to the cycle-matching and spectral conditions stated above. Detailed connections between recurrence witnesses and purity-loss estimation are developed in ~\cite{pandit2022bounds,penshin2024robust}.

These considerations suggest a practical benchmarking workflow. DFE is well suited for small- to intermediate-size demonstrations on structured target states and for validating software pipelines. The survival-probability method may be useful where repeated return experiments are accessible and a compact adjoint-cycle proxy is informative, provided that the cycle-matching and spectral assumptions hold. Its relative performance under readout noise and other SPAM errors remains to be established experimentally.

\subsection{Outlook}
Several extensions naturally emerge from this study. One important direction is fidelity estimation for target states that are classically difficult to simulate, including highly magical nonstabilizer states. For a pure $N$-qubit state $\rho$, a useful measure of nonstabilizerness is the order-2 stabilizer Rényi entropy~\cite{leone2022stabilizer},
\begin{equation}
M_2(\rho)
=-\log_2\!\left[
\frac{1}{2^N}
\sum_{P\in\mathcal{P}_N}
\left|\Tr(\rho P)\right|^4
\right],
\label{eq:stabilizer-renyi-entropy}
\end{equation}
where $\mathcal{P}_N=\{I,X,Y,Z\}^{\otimes N}$ denotes the set of $4^N$ unphased Pauli strings. Equation~\eqref{eq:stabilizer-renyi-entropy} is stated for pure states; mixed states require a suitably normalized generalization. Our numerical observations indicate that evaluating the target-state Pauli structure becomes increasingly costly as the state magic increases. Indeed, the sampling resources required by generic DFE protocols can grow exponentially with the stabilizer Rényi entropy~\cite{leone2023hardness}. The classical cost of characterizing the target therefore becomes part of the benchmarking problem itself, and determining how this cost scales with state magic remains an important open question.

A second direction is to reduce the DFE sampling overhead through more advanced statistical procedures. Grouped-Pauli sampling~\cite{barbera2025sampling}, adaptive measurement-budget allocation, and Markov chain Monte Carlo ideas may all provide practical improvements in regimes where naive importance sampling remains costly.

A third direction concerns the survival-probability framework. Deriving the coefficient optimization directly from a noise model and analyzing error propagation under finite-shot fluctuations and imperfect inverses would make the protocol more directly comparable to standard randomized benchmarking and related scalable characterization methods~\cite{magesan2012scalable}.

\section{Conclusion}
We have presented and analyzed direct fidelity estimation (DFE) and a survival-probability-based reconstruction of an adjoint-cycle fidelity proxy. These approaches offer alternatives to full quantum-state tomography, subject to their respective requirements: DFE relies on accessible target-state Pauli information, while the survival-probability reconstruction requires a cycle matching $\mathcal{Q}=\mathcal{K}^{\dagger}\mathcal{K}$ and a spectrum within the approximation interval. These qualifications matter when assessing their practical use on near-term quantum devices.

Direct fidelity estimation offers an efficient and statistically rigorous route to fidelity estimation when the Pauli structure of the target state is known and can be sampled effectively. Our analysis demonstrates that attenuation-aware corrections can substantially improve robustness against measurement noise, extending the practical applicability of DFE under realistic experimental conditions. In parallel, the survival-probability method provides a complementary reconstruction of the adjoint-cycle proxy $F_{\mathcal{Q}}$ from return-probability moments when the implemented cycle equals $\mathcal{Q}$ and its spectrum meets the approximation assumptions. The $L_2$-optimized coefficients significantly outperform standard Taylor-based approximations in the simulated moderate- and strong-noise regimes. Because these simulations use exact moments, they do not establish robustness to finite-shot or explicit SPAM errors.

Taken together, these methods highlight two distinct yet synergistic philosophies of quantum benchmarking: one based on structural information encoded in the target state, and the other based on dynamical information encoded in experimentally observable recurrences. The framework presented here therefore provides a flexible toolbox for fidelity estimation across a broad range of experimental platforms and noise environments.

Looking ahead, an important next step is the deployment of these protocols on real superconducting quantum hardware. In particular, the recently upgraded quantum-computing infrastructure at VTT provides an ideal testbed for studying the practical performance of fidelity estimators beyond numerical simulation. Future work will focus on implementing DFE and survival-probability protocols directly on experimentally prepared multiqubit states, comparing their performance against established characterization techniques such as randomized benchmarking, cycle benchmarking, and scalable noise-learning frameworks based on sparse Pauli--Lindblad models. Particular emphasis will be placed on quantifying their resilience to calibration drifts, leakage, correlated noise, coherent control errors, and readout imperfections, while assessing their suitability for integration into automated validation and calibration workflows for fault-tolerant quantum processors ~\cite{korkiakoski2024flagship,carugno2024nordiquest,magesan2012scalable,erhard2019characterizing,proctor2022scalable,senko2023realization}.

Beyond benchmarking individual states, the long-term vision is to integrate scalable fidelity estimation into automated quantum-device validation pipelines. Such tools could enable continuous monitoring of processor performance, rapid diagnosis of noise mechanisms, and data-driven optimization of quantum control strategies. As quantum processors continue to scale toward error-corrected operation, robust and resource-efficient fidelity estimation will become an increasingly important component of the experimental quantum-computing stack, bridging the gap between device characterization and noise diagnosis~\cite{penshin2024robust}, control optimization~\cite{da2026geodesic}, and fault-tolerant quantum computation~\cite{devitt2013quantum,terhal2015quantum,campbell2017roads,acharya2023suppressing,korkiakoski2024flagship}

\appendix
\section{Pulse-Inverse Construction and Spectral Conditions}
\label{app:survival_derivations}%
\subsection{Magnus expansion and the physical pulse inverse}
\label{app:pulse_inverse}%

For time-dependent Markovian evolution,
\begin{equation}
\frac{d}{dt}\lvert\rho(t)\rangle\rangle
=\left[-i\mathcal{H}(t)+\mathcal{L}(t)\right]
\lvert\rho(t)\rangle\rangle.
\end{equation}
In the interaction picture defined by
$\lvert\rho_{\mathrm{int}}(t)\rangle\rangle
=\mathcal{U}^{\dagger}(t)\lvert\rho(t)\rangle\rangle$,
the dissipator becomes
\begin{equation}
\mathcal{L}_{\mathrm{int}}(t)
=\mathcal{U}^{\dagger}(t)\mathcal{L}(t)\mathcal{U}(t),
\end{equation}
and the physical channel can be written using the Magnus expansion~\cite{blanes2009magnus}:
\begin{equation}
\mathcal{K}(T)=\mathcal{U}(T)
\exp\!\left(\Omega_1+\Omega_2+\cdots\right),
\label{eq:app_magnus_def}
\end{equation}
where
\begin{align}
\Omega_1&=\int_0^T\mathcal{L}_{\mathrm{int}}(t)\,dt,\\
\Omega_2&=\frac{1}{2}\int_0^T\!dt_1\int_0^{t_1}\!dt_2
\,[\mathcal{L}_{\mathrm{int}}(t_1),
\mathcal{L}_{\mathrm{int}}(t_2)].
\end{align}

A pulse inverse reverses the control Hamiltonian in amplitude and time order, $H_I(t)=-H(T-t)$, but does not reverse the sign of the environmental dissipator; instead, $\mathcal{L}_I(t)=\mathcal{L}(T-t)$. Neglecting inverse-control errors and higher Magnus orders gives
\begin{equation}
\mathcal{K}_I(T)\simeq
e^{\Omega_1^{(I)}}\mathcal{U}^{\dagger}(T).
\label{eq:app_ki_def}
\end{equation}
The forward--inverse cycle is therefore
\begin{align}
\mathcal{C}=\mathcal{K}_I\mathcal{K}
&\simeq e^{\Omega_1^{(I)}}e^{\Omega_1}\\
&=\exp\!\left[
\Omega_1^{(I)}+\Omega_1
+\frac{1}{2}[\Omega_1^{(I)},\Omega_1]+\cdots
\right].
\label{eq:app_cycle_bch}
\end{align}
Only when $\Omega_1^{(I)}\simeq\Omega_1$ and the commutator corrections are negligible does this reduce to
\begin{equation}
\mathcal{C}\simeq e^{2\Omega_1}.
\label{eq:app_kik_doubling}
\end{equation}
Under the same assumptions, the leading purity losses satisfy
\begin{align}
\Delta P_{\mathcal{K}}
&=-2\operatorname{Re}
\langle\langle\rho_0\vert\Omega_1\vert\rho_0\rangle\rangle,\\
\Delta P_{\mathcal{C}}
&=-4\operatorname{Re}
\langle\langle\rho_0\vert\Omega_1\vert\rho_0\rangle\rangle
=2\Delta P_{\mathcal{K}},
\label{eq:app_kik_purity_loss}
\end{align}
up to $\mathcal{O}(\lVert\Omega_1\rVert^2,\lVert\Omega_2\rVert)$ corrections. Coherent errors in the implemented inverse introduce additional generators and limit this cancellation; pulse-inverse constructions and their error assumptions are discussed in ~\cite{santos2023scalable}.

\subsection{Spectral interval and domain of validity}
\label{app:spectral_bounds}%

The physical pulse-inverse cycle $\mathcal{C}=\mathcal{K}_I\mathcal{K}$ should be distinguished from the positive adjoint cycle
\begin{equation}
\mathcal{Q}=\mathcal{K}^{\dagger}\mathcal{K}.
\end{equation}
The latter is Hermitian and positive semidefinite, with eigenvalues equal to the squared singular values of $\mathcal{K}$. To bound its smallest eigenvalue, define the interaction-picture propagator
\begin{equation}
\mathcal{E}(t)=\mathcal{U}^{\dagger}(t)\mathcal{K}(t),
\qquad
\dot{\mathcal{E}}(t)=\mathcal{L}_{\mathrm{int}}(t)\mathcal{E}(t).
\end{equation}
Its inverse obeys
\begin{equation}
\frac{d}{dt}\mathcal{E}^{-1}(t)
=-\mathcal{E}^{-1}(t)\mathcal{L}_{\mathrm{int}}(t),
\label{eq:app_inverse_diff}
\end{equation}
where the operator ordering is important. Submultiplicativity and Gronwall's inequality give
\begin{equation}
\lVert\mathcal{E}^{-1}(T)\rVert
\leq e^{\eta},
\qquad
\eta=\int_0^T\lVert\mathcal{L}(t)\rVert\,dt,
\label{eq:app_inverse_bound}
\end{equation}
where unitary invariance of the norm has been used. Since $\mathcal{K}=\mathcal{U}\mathcal{E}$,
\begin{equation}
s_{\min}(\mathcal{K})
=\frac{1}{\lVert\mathcal{K}^{-1}\rVert}
\geq e^{-\eta}.
\label{eq:app_smin_bound}
\end{equation}
It follows that
\begin{equation}
\lambda_{\min}(\mathcal{Q})
=s_{\min}^2(\mathcal{K})
\geq e^{-2\eta}.
\label{eq:app_min_lambda_bound}
\end{equation}
Thus, if $\lambda_{\max}(\mathcal{Q})\leq1$, the polynomial interval in \cref{eq:l2_error} may be chosen with $g\leq e^{-2\eta}$. Because Hilbert--Schmidt contractivity is not automatic for every quantum channel, the upper bound must either be verified for the channel under study or enforced by rescaling the spectrum. Moreover, moments of the experimentally implemented cycle $\mathcal{C}$ coincide with moments of $\mathcal{Q}$ only when $\mathcal{K}_I=\mathcal{K}^{\dagger}$.

Finally, the weak-noise condition is
\begin{equation}
\eta=\int_0^T\lVert\mathcal{L}(t)\rVert\,dt\ll1,
\label{eq:app_validity_condition}
\end{equation}
together with $\lVert\Omega_2+\Omega_3+\cdots\rVert\ll
\lVert\Omega_1\rVert$ when the Magnus series is truncated at first order. Equivalently, closeness of the adjoint-cycle spectrum to unity can be expressed as $e^{2\eta}-1\ll1$; the condition $e^{2\eta}\ll1$ would be impossible for $\eta\geq0$.

\section{Depolarizing-Channel Fidelity Derivation}
\label{app:depolarizing-fidelity}

For an arbitrary operator $X$, the linear extension of the depolarizing channel is
\begin{equation}
\Lambda_p(X)=(1-p)X+\frac{p}{d}\Tr(X)\mathbb{I}.
\label{eq:app-depolarizing-linear}
\end{equation}
It is trace-preserving, and for any unitary $V$ it is covariant:
\begin{equation}
\Lambda_p(VXV^\dagger)=V\Lambda_p(X)V^\dagger,
\end{equation}
as follows from invariance of the trace under unitary conjugation. Thus, for a pure state $\rho=\ketbra{\psi}{\psi}$, the noisy state is $\sigma=(1-p)\rho+p\mathbb{I}/d$. The pure-state reduction of Uhlmann fidelity and $\Tr(\rho^2)=\Tr(\rho)=1$ give
\begin{align}
F(\rho,\sigma)
&=\bra{\psi}\sigma\ket{\psi}\\
&=(1-p)+\frac{p}{d}
=1-p\left(1-\frac{1}{d}\right),
\end{align}
which is independent of the particular pure state and of any preceding ideal unitary.

\section*{Data Availability}
The data and code that support the findings of this study are available at GitHub \href{https://github.com/gimdall}{SP and DFE}.

\begin{acknowledgments}
T.P. acknowledges financial support from the Strategic Council of Finland through the Finnish Quantum Flagship project (Grant No. 359284) and thanks Raam Uzdin for stimulating discussions and valuable insights during his Ph.D. studies. K.C. acknowledges support from the Danish National Research Foundation through the Center for Macroscopic Quantum States (bigQ, DNRF142), the EU project CLUSTEC (Grant Agreement No. 101080173), and the ERC Advanced Grant ClusterQ (Grant Agreement No. 101055224, ERC-2021-ADG).
\end{acknowledgments}
\bibliographystyle{apsrev4-2}
\bibliography{references}

\end{document}